\documentclass[twoside,11pt]{article}
\usepackage{jair, theapa, rawfonts}

\jairheading{77}{2023}{459-485}{09/2022}{06/2023}
\ShortHeadings{The Complexity of Matching Games: A Survey}
{Benedek, Bir\'o, Johnson, Paulusma, \& Ye}
\firstpageno{459}

\usepackage{theapa}
\usepackage{amssymb}
\usepackage[utf8]{inputenc}
\usepackage[small]{caption}
\usepackage{lineno,soul,graphicx,amsmath,booktabs,tikz,xcolor}
\usetikzlibrary{arrows, automata,shapes}
\usetikzlibrary{decorations.pathreplacing}

\newtheorem{open}{Open Problem}

\newtheorem{theorem}{Theorem}

\newcommand{\R}{\mathbb{R}}
\newcommand{\NP}{{\sf NP}}
\newcommand{\vd}{{\tt vd}}
\newcommand{\va}{{\tt va}}
\newcommand{\ed}{{\tt ed}}

\newcommand{\ea}{{\tt ea}}
\newcommand \dia{\hfill{$\diamond$}}

\begin{document}

\title{The Complexity of Matching Games: A Survey}

\author{\name M\'arton Benedek \email benedek.marton@krtk.hu \\
       \name P\'eter Bir\'o \email biro.peter@krtk.hu \\
       \addr Institute of Economics, KRTK,\\
       T\'oth K\'alm\'an u. 4., Budapest, Hungary\\
       Corvinus University of Budapest\\
       F\H{o}v\'am t\'er 8., Budapest, Hungary
       \AND
       \name Matthew Johnson \email matthew.johnson2@durham.ac.uk \\
       \name Dani\"el Paulusma \email daniel.paulusma@durham.ac.uk\\
       \name Xin Ye \email xin.ye@durham.ac.uk\\
       \addr Department of Computer Science, Durham University,\\
       Upper Mountjoy Campus, Stockton Road, Durham, UK}

\maketitle

\begin{abstract}
Matching games naturally generalize assignment games, a well-known class of cooperative games.
Interest in matching games has grown recently due to some  breakthrough results and new applications.
This state-of-the-art survey provides an overview of matching games and
extensions, such as $b$-matching games and
partitioned matching games; the latter originating from the emerging area of international kidney exchange.
In this survey we focus on computational complexity aspects of various
game-theoretical solution concepts, such as the core, nucleolus and Shapley value, when the input is restricted to a
{matching game or one of its variants.}
\end{abstract}

\section{Introduction}\label{s-intro}

\noindent
We consider a wide range of matching problems that involve {monetary} payments. 
In order to illustrate this, we start with an artificial but simple example {where the payments come in the form of prize money. This example} has its origins in a paper of~\citeA{EK01}.

\subsection{An Initial Example}\label{s-1}

\noindent
Suppose a group~$N$ of tennis players play a doubles tennis tournament. Each pair of players estimates the expected prize money when they form a doubles partnership. 
{Two players that are not matched to each other may leave their partner if together they can win more prize money.}  \citeA{EK01} studied the question of whether the players can be {\it matched} in pairs {in such a way} that {this situation does not occur}.
A wider question is whether the players can be matched together such that no {\it group} of players {expects to be awarded more prize money} by playing an event on their own. This situation describes a {\it matching game}; see Section~\ref{s-mg} for a formal definition.

\subsection{A  Second Example: International Kidney Exchange}\label{s-2}

\noindent
Our next example is more elaborate and leads to the notion of a {\it partitioned matching game}, which we also formally define in Section~\ref{s-mg}. This example is related to an emerging application area, namely that of  {\it international kidney exchange}.
This application area formed the main motivation for us to write our survey and below we discuss it in detail.

Chronic kidney disease can lead to kidney failure, which can have a devastating impact on patients' lives. Currently, the most effective treatment for kidney failure is transplantation, which could be either from a deceased or living donor. The long-term outcomes of the latter type of donation are best. However, one major obstacle that prevents a living donor from donating a kidney to a family member or friend is medical incompatibility between patient and donor. Namely, with high probability the donor's kidney would be rejected by the patient's body.

\begin{figure}
\begin{center}
\begin{tabular}{cc}
 \begin{tikzpicture}[
            > = stealth,
            shorten > = 1pt,
            auto,
            node distance = 3cm,
            semithick
        ]

        \tikzstyle{every state}=[scale=0.75,
            draw = black,
            thick,
            fill = white,
            minimum size = 4mm
        ]
        \node[state] (i2) {$i_2$};
        \node[state] (i1) [above right of=i2] {$i_1$};
        \node[state] (j2) [right of=i2] {$j_2$};
        \node[state] (j1) [below right of=i2] {$j_1$};
        \node[state] (i3) [right of=j2] {$i_3$};
 \path[->]  (i1) edge node {} (i2);
        \path[->] [bend left] (i2) edge node {} (i1);
        \path[->]  (i2) edge node {} (j2);
        \path[->] [bend left] (j2) edge node {} (i2);
        \path[->] (i2) [bend right] edge node {} (j1);
        \path[->] (j2) edge node {} (i1);
                \path[->] (j2) [bend left] edge node {} (j1);
        \path[->] (j1) edge node {} (j2);
        \path[->] (i1) edge node {} (i3);
    \end{tikzpicture}
    \hspace*{2.5cm}
     \begin{tikzpicture}[
            > = stealth,
            shorten > = 1pt,
            auto,
            node distance = 3cm,
            semithick
        ]

        \tikzstyle{every state}=[scale=0.75,
            draw = black,
            thick,
            fill = white,
            minimum size = 4mm
        ]

        \node[state] (i2) {$i_2$};
        \node[state] (i1) [above right of=i2] {$i_1$};
        \node[state] (j2) [right of=i2] {$j_2$};
        \node[state] (j1) [below right of=i2] {$j_1$};
        \node[state] (i3) [right of=j2] {$i_3$};
 \path[]  (i1) edge node {} (i2);
        \path[]  (i2) edge node {} (j2);
        \path[] (j1) edge node {} (j2);
    \end{tikzpicture}
\end{tabular}
\end{center}
\caption{{\it Left:} a pool of patient-donor pairs, where an arc from a pair $i=(p,d)$ to a pair $j=(p',d')$ indicates that donor $d$ is compatible with patient $p'$. {\it Right:} the corresponding compatibility graph.}\label{f-comp}
\end{figure}
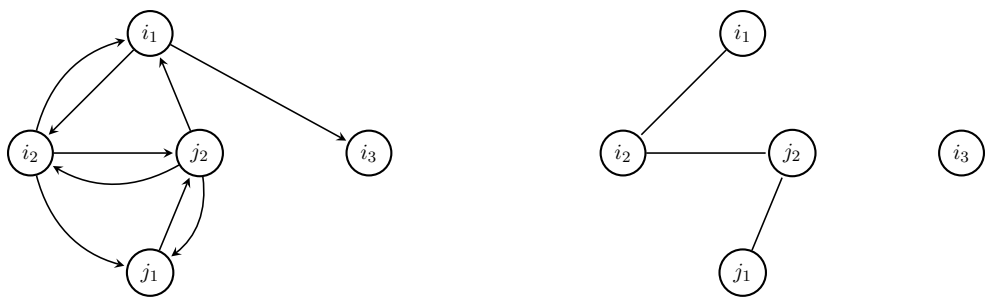

To circumvent medical incompatibility, national Kidney Exchange Programmes (KEPs) have been established in many countries
in the last 30 years~\shortcite{Bi_etal2019}. In a KEP, a patient and their willing but incompatible donor are placed in a pool with other patient-donor pairs. 
Now, if donor~$d$ of patient~$p$ is compatible with patient~$p'$ and simultaneously, donor~$d'$ of $p'$ is compatible with $p$, then {a {\it pairwise exchange} can be made, that is,} $d$ and~$d'$ can be ``swapped''. In this way, both $p$ and $p'$ can now receive treatment. Hence, the problem is to find a solution in a given pool of patient-donor pairs that helps as many patients as possible.
Such a solution corresponds to a {maximum} matching in the {\it compatibility graph}, which is obtained by introducing a vertex for each patient-donor pair in the pool and adding an edge between two vertices $(p,d)$ and $(p',d')$ if and only if $d$ is compatible with $p'$ and $d'$ is compatible with $p$; see Figure~\ref{f-comp} for an
example.\footnote{Countries may also allow cycles, e.g. $d_1$ donates to $p_2$, $d_2$ to $p_3$ and $d_3$ to $p_1$; or paths,  e.g.
an altruistic donor $d^*$ donates to $p_1$, $d_1$ to $p_2$, $d_2$ to $p_3$ and $d_3$ to the
deceased donor waiting list. More patients now receive treatment. {However,} in many countries altruistic donors are forbidden. {Moreover}, cycles must be kept small to prevent them from breaking. {Therefore, in several countries only pairwise exchanges are allowed, such as in France and Hungary~\shortcite{Bi_etal2019}. If the upper bound on the cycle length is larger than~$2$ or if altruistic donors are allowed, then we can no longer model the problem of maximizing the number of transplants as a matching problem; the problem also becomes \NP-hard \shortcite{ABS07}.}}

KEPs operate in regular rounds, each time considering an updated pool, where some patient-{donor} pairs have left and new ones have come in. {In each round, the primary goal of a KEP is to maximize the number of kidney transplants. However, KEPs may also take other factors into account, such as the quality of the transplants. Transplant qualities
can be represented by assigning weights to the edges of the compatibility graph. The goal is then to find a matching of maximum weight in the compatibility graph.}

As merging pools of national KEPs could lead to even more patients being
treated, {\it international} KEPs are now starting to be formed.  The problem now is not just to treat many patients, but to do so in such a way that each country believes they are being treated fairly. 
{If some countries receive fewer kidneys than they could receive by forming a smaller international KEP or by running their own KEP again, then the international KEP may fall apart.}
Thus we must focus on the \emph{stability} of international KEPs.

\shortciteA{KNPV20} introduced a credit system to ensure stability of international KEPs. In their model, the number of kidney transplants for a country~$i$ is specified by a ``fair'' target number for~$i$. The differences between the actual number of transplants for $i$ and this target number is given as credits for~$i$ in the next round. Simulations using their model have also been performed, with different parameters, by \shortciteA{BGKPPV20}, and 
 {for international KEPs with pairwise exchanges only, but with a larger number of countries than the two previous studies,} by \shortciteA{BBKP22}.  One of the central considerations in this model is to decide what target numbers are fair. 
{\shortciteA{BKPP19} modelled an international KEP with pairwise exchanges only}
as a partitioned matching game, in which each country is a player. {This makes it possible to} use solution concepts from Cooperative Game Theory for finding fair target allocations. {W}e will explain {this} in more detail in Section~\ref{s-mg}.

\subsection{Overview of the Survey}

\medskip
\noindent
In Section~\ref{s-background} we introduce some background information on
Cooperative Game Theory and Graph Theory.
This enables us to formally define the concept of a matching game, its special case of an assignment game and two closely related generalizations of matching games in Section~\ref{s-mg}, one of which is the notion of a partitioned matching game and the other is
known as the $b$-matching game. In the same section, we also define the
fractional matching game, a known relaxation of a matching game.
In Section~\ref{s-nbg} we discuss the well-known network bargaining games,
which have a close relationship with matching games.

In Sections~\ref{s-co}--\ref{s-sv} we survey known computational complexity results of the core, nucleolus and Shapley value, respectively, for assignment games, $b$-assignment games, matching games, $b$-matching games and partitioned matching games. Unless specified otherwise, running times of algorithms are expressed in the input size $n+m$ of the underlying (weighted) graph $G=(V,E)$ where $n=|V|$ and $m=|E|$.
In Section~\ref{s-co} we explain in particular to what extent solutions for network bargaining games correspond to finding core allocations of matching games and their variants. In the same section we also present complexity results on {\it core stabilizers}. Stabilizers allow for making small modifications to the underlying graph, such that the adjusted game has a non-empty core.

Throughout our survey we list relevant open problems, and in Section~\ref{s-con} we conclude our survey by listing some more.

\section{Background}\label{s-background}

\medskip
\noindent
In Section~\ref{s-cgt} we give some basic {cooperative} game theory terminology and notation; {see also the textbooks of \citeA{Ow13} and \citeA{CEW11}.}
In Section~\ref{s-graph} we give the necessary graph theory terminology and notation; 
{see also the textbooks of \citeA{Di12} and \citeA{BM08}.}

\subsection{Cooperative Game Theory}\label{s-cgt}

\medskip
\noindent
A \emph{(cooperative) game} is a pair $(N,v)$, where {$N=\{1,\dots,|N|\}$} is a set of \emph{players} (agents)
and $v: 2^N\to \R_+$ is a \emph{value function} {(or \emph{chacteristic function})} with $v(\emptyset) = 0$; {{here, $R_+$ denotes the set of non-negative real numbers.} A {\it coalition} is a subset of~$N$.
For a coalition $S\subseteq N$, the value $v(S)$ describes the joint profit or costs if the players in $S$ collaborate with each other. In line with the type of game we consider in this survey, we let $v$ be a profit function in the remainder of the paper. Then, assuming that $v(N)$ is maximum over all partitions of $N$, the central problem is how to keep the {\it grand coalition} $N$ stable by distributing the {\it total value} $v(N)$ amongst the players of $N$ in a fair way.

An {\it allocation} {(or {\it pre-imputation})} for a game $(N,v)$ is a vector $x \in \R^N$ with $x(N) = v(N)${;} here, for $S\subseteq N$, we write $x(S)=\sum_{i\in S}x_i$.
A {\it solution concept} prescribes a set of allocations for every game.
Each solution concept has its own fairness properties and computational complexities;
the choice of a certain solution concept depends on context. In this survey we focus on computational complexity aspects {of solution concepts}.

The {\it core} {was formally introduced by \citeA{Gi59} and} is the best-known solution concept.
It consists of all allocations $x \in \R^N$ with $x(S)\geq v(S)$ for every $S\subseteq N$.
Core allocations {(that is, allocations that belong to the core)} ensure stability, as no group of players has any incentive to leave {the grand coalition}~$N$.
However, the core of a game might be empty. In Section~\ref{s-mg} we illustrate by an example that this may happen even for matching games with only three players.

Should the core be empty we might seek for an allocation $x$ that satisfies all core constraints ``as much as possible'' by solving the following linear program:
\begin{eqnarray*}
{(\mbox{LC})} &\varepsilon_1 :=& \max \varepsilon\\
                     && x(S) \ge v(S) + \varepsilon \quad \mbox{for all}\; S\in 2^N\backslash\{\emptyset,N\}\\
                      && x(N) = v(N).
\end{eqnarray*}

\noindent
The set of allocations $x$ for which $(x,\varepsilon_1)$ is an
optimum solution of $(\mbox{LC})$ {has been introduced by \shortciteA{MPS79} as} the \emph{least core} of $(N,v)$. The least core belongs to the core if the core is non-empty, which is the case if and only if $\varepsilon_1\geq 0$. {Note that by excluding the sets $\emptyset$ and $N$ in the definition of (LC), the least core might be a proper subset of the core, should the core be non-empty.}

In contrast to the core, the {\it Shapley value} $\phi(N,v)$ always exists.
It is the unique singleton solution concept that satisfies some set of desirable fairness properties~\cite{Sh53}.
It assigns every player $i\in N$:
\begin{equation*}\label{eq-Shapley}
\phi_i(N,v) = \displaystyle\sum_{S \subseteq N\backslash \{i\}}
 \frac{|S|!({|N|}-|S|-1)!}{{|N|}!}\bigg(v(S\cup \{i\})-v(S)\bigg).
\end{equation*}
{The Shapley value might be negative but not for the games we consider.
Moreover}, the Shapley value might not be in the core even if the core is non-empty. This may happen even for matching games, as we will
demonstrate in Section~\ref{s-thecore} with an example.

We now define another well-known solution concept, namely the nucleolus. In contrast to the Shapley value, the nucleolus is always in the core if the core is non-empty.
The {\it excess} of a non-empty coalition $S \subsetneq N$ for an allocation $x$ is given by $e(S,x) = x(S)-v(S)$.
Ordering the $2^{{|N|}}-2$ excesses non-decreasingly gives us the {\it excess vector} $e(x)$.
The {\it nucleolus} $\eta(N,v)$ is the unique allocation~\cite{Sc69} that lexicographically maximizes $e(x)$ over the set of {\it imputations} $I(N,v)$, which consists of all allocations~$x$ with $x_i\geq v(\{i\})$ for $i\in N$; the nucleolus is only defined if $I(N,v)$ is non-empty.
In Sections~\ref{s-mg} and~\ref{s-thecore} we give some examples of matching games, for which we computed the nucleolus.

An alternative definition of the nucleolus is given by
\shortciteA{MPS79}; solve for $r\geq 1$, the following sequence of linear programs until a unique solution, which gives us the nucleolus, remains:
\begin{eqnarray*}
{(\mbox{LP}_r)}\hspace*{-2mm} &\varepsilon_r := \hspace*{-2mm}& \max \varepsilon\\
                      && x(S) \ge v(S) + \varepsilon \quad {\mbox{for all}}\; S\in 2^N\backslash\{ {\cal S}_0\cup\ldots\cup {\cal S}_{r-1}\}\\
                      && x(S) = v(S) + \varepsilon_i \;\:\; {\mbox{for all}}\; S\in {\cal S}_i \quad (0\leq i\leq r-1)\\
                      && x\in I(N,v),
\end{eqnarray*}
where $\varepsilon_0=-\infty$, ${\cal S}_0=\emptyset$ and for $i\geq 1$, ${\cal S}_i$ consists of all coalitions~$S$ with  $x(S) = v(S) + \varepsilon_i$ for every optimal solution $(x,\varepsilon_{i})$ of $(\mbox{LP}_{i})$.
From this definition {and the definition of the least core}, it follows that {if the core is non-empty, then the nucleolus exists and even belongs to the least core}.

{Finally, we define the pre-kernel, introduced by \shortciteA{MPS71}. We let $$s_{ij}(x)=  \min\{e(S,x)\; |\; S\subseteq N\; \mbox{with}\; i\in S, j\notin S\}.$$
The {\it pre-kernel} of a game $(N,v)$ is the set of all allocations $x\in \R^N$ with $s_{ij}(x)=s_{ji}(x)$ for every pair of distinct players $i,j\in N$.}

\subsection{Graph Theory}\label{s-graph}

\medskip
\noindent
Throughout this section, the graph $G$ is an undirected graph with no self-loops and no multiple edges. Let $V$ be the vertex set of $G$, and let $E$ be the edge set of $G$. 
{We denote an edge between two vertices $i$ and $j$ as $ij$ (or $ji$).} 
For a subset $S\subseteq V$, we let $G[S]$ denote the subgraph of $G$ {\it induced} by $S$, that is, $G[S]$ is the graph obtained from $G$ after deleting all vertices that are not in $S$ {(and their incident edges)}. We say that $G$ is {\it bipartite} if there exist two disjoint vertex subsets~$A$ and~$B$ such that $V=A\cup B$ and every edge in $E$ joins a vertex in  $A$ to a vertex in~$B$.

 A function $f$ that maps some set $X$ to $\R$ is positive if $f(x)\; {>}\; 0$ for every $x\in X$. 
 {For an integer~$t$, we write $f\equiv t$ if $f(x)=t$ for every $x\in X$, and we write $f\leq t$ if $f(x)\leq t$ for every $x\in X$.}
 
Now, let $M\subseteq E$ be some subset of edges.
For a positive edge weighting~$w$ of $G$, the {\it weight} of $M$ is defined as $w(M)=\sum_{e\in M}w(e)$.
For a positive vertex capacity function $b$, we say that $M$ is a {\it $b$-matching} of $G$ if each~$i \in V$ is incident to at most
$b(i)$ edges in $M$. If $b\equiv 1$, then $M$ is a {\it matching}.

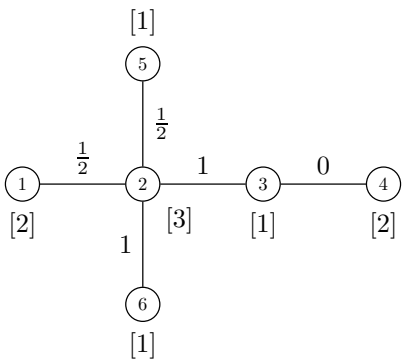
\begin{figure}[t]
\centering
\begin{tikzpicture}[scale=0.8,rotate=0,line width=0.48]
\draw
(0, 0) node[label=below:{\small[$2$]},circle, black, scale=0.7,draw](a){\small $1$}
(2, 0) node[label=below right:{\small[$3$]},circle, black, scale=0.7,draw](b){\small $2$}
(4, 0) node[label=below:{\small[$1$]},circle, black, scale=0.7,draw](c){\small $3$}
(6, 0) node[label=below:{\small[$2$]},circle, black, scale=0.7,draw](d){\small $4$}
(2, 2) node[label=above:{\small[$1$]},circle, black, scale=0.7,draw](e){\small $5$}
(2,-2) node[label=below:{\small[$1$]},circle, black, scale=0.7,draw](f){\small $6$};
\draw[-] (a) -- node[above] {\small $\frac{1}{2}$}(b);
\draw[-] (b)--node[above] {\small $1$}(c);
\draw[-] (c)--node[above] {\small $0$}(d);
\draw[-] (e) --node[right] {\small $\frac{1}{2}$}(b);
\draw[-] (b) --node[left] {\small $1$}(f);
\end{tikzpicture}
\caption{An example of a half-$b$-matching $f$ in a graph $G$ with $w\equiv 1$ and where~$b$ is given by $b(3)=b(5)=b(6)=1$; $b(1)=b(4)=2$; and
$b(2)=3$, {as illustrated by the $[x]$ labels in the figure.} Note that {$f(12)=f(25)=\frac{1}{2}$}; {$f(23)=f(26)=1$};  {$f(34)=0$}; and $w(f)=3$.}\label{f-half}
\end{figure}

A \emph{fractional $b$-matching} of $G$ is an edge mapping $f:E\rightarrow [0,1]$ such that $\sum_{e:i\in e}f(e)\leq b(i)$ for each $i\in {V}$. The \emph{weight} of a fractional $b$-matching $f$ is defined as $w(f)=\sum_{e\in E}w(e)f(e)$. If $b\equiv 1$, then $f$ is a {\it fractional matching}. Note that $f$ is a $b$-matching of $G$ if and only if $f(e)\in \{0,1\}$ for every $e\in E$. We say that $f$ is a \emph{half-$b$-matching}  of $G$ if  $f(e)\in \{0,\frac{1}{2},1\}$ for every $e\in E$. If $b \equiv 1$, then a half-$b$-matching of $G$ is also called a {\it half-matching}. See Figure~\ref{f-half} for an example of a half-$b$-matching.

\section{Matching Games and Generalizations}\label{s-mg}

\medskip
\noindent
In this section, we define matching games, assignment games, $b$-matching games, $b$-assignment games and partitioned matching games. {We also} show how the two examples from Section~\ref{s-intro} can be seen as matching games and partitioned matching games, respectively.

\bigskip
\noindent
{\bf Definition 1.} 
{\shortcite{DIN99}}
 A {\it matching game} on a graph~$G=(V,E)$ with a positive edge weighting~$w$ is the game $(N,v)$
where $N=V$ and for $S\subseteq N$, the value $v(S)$ is the maximum weight~$w(M)$ over all matchings $M$ of $G[S]$.

\bigskip
\noindent
Matching games on bipartite graphs are commonly known as {\it assignment games} {\cite{SS72}}.  In the example of Section~\ref{s-1}, the tennis players form the vertices of a graph such that there exists an edge between two vertices $i$ and $j$ if and only if $i$ and $j$ can form a doubles pair together, with expected prize money $w(ij)$.
A game on a graph $G$ with edge weighting $w$ is {\it uniform} (or {\it simple}) if $w\equiv 1$.

\begin{figure}[b]
\centering
\begin{tikzpicture}[scale=0.6,rotate=0,line width=0.48]
\draw
(4, 0) node[circle, black, scale=0.6,draw](c){h}
(6, 0) node[circle, black, scale=0.6,draw](d){i}
(5, 2) node[circle, black, scale=0.6,draw](e){j};

\draw[-] (c) --(d) --(e)--(c);
\end{tikzpicture}
\caption{The graph ${K_3}$, whose corresponding uniform matching game has an empty core.}\label{C_3}
\end{figure}
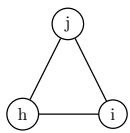

Already uniform matching games may have an empty core. For example, consider the uniform matching game $(N,v)$ defined on a graph $G$ that is a triangle with vertices $h,i,j$ and edges $hi$, $ij$ and $jh$; see also Figure~\ref{C_3}. The reason for core-emptiness for matching games is that every
imputation~$x$ has a so-called blocking pair. That is, a pair of adjacent players $\{i,j\}$ in a matching game $(N,v)$ is a {\it blocking pair} for an imputation $x$ if $x_i+x_j<w(ij)$ holds. The {\it blocking value} of $\{i,j\}$ for $x$ is defined as
$$e_x(ij)^+=\max\{0,w(ij)-(x_i+x_j)\}.$$
\noindent
 If $\{i,j\}$ is not blocking $x$, then $e_x(ij)^+=0$; otherwise,
$e_x(ij)^+$ expresses the extent to which $\{i,j\}$ blocks $x$. We let
$B(x)=\sum_{ij\in E}e_x(ij)^+$ be the {\it total blocking value} for $x$.
{\shortciteA{DIN99} showed that} an imputation $x$ is in the core of a matching game $(N,v)$ defined on a graph $G$ with positive edge weighting $w$ if and only {if} for every $ij\in E$,
$$x_i+x_j\geq w(ij),$$
\noindent
so if and only if $B(x)=0$ and $x$ has no blocking pairs.
{This follows as if there is a blocking pair then $x$ is not in the core, and, for the other direction, suppose that $S$ is a blocking coalition. 
Let $M$ be a maximum weight matching in $G[S]$.
Then $x(S)<v(S)=w(M)=\sum_{e_i\in M}w(e_i)$, and so for at least one of the edges $ij\in M$, $x_i+x_j<w(ij)$ must hold; hence $x$ not in the core.}
For a (larger) example of a matching game with a non-empty core, we refer to Example~2 in Section~\ref{s-thecore}.

\bigskip
\noindent
{\bf Definition 2.} {\shortcite{BKPW18}}
A {\it $b$-matching game} on a graph $G=(V,E)$ with a positive vertex capacity function $b$ and a positive edge weighting~$w$ is the game $(N,v)$ where $N=V$
and for $S\subseteq N$, the value $v(S)$  is the maximum weight $w(M)$ over all $b$-matchings~$M$ of $G[S]$.

\bigskip
\noindent
We refer to {Figure~\ref{f-fourth}} for an example. If $b\equiv 1$, then we have a matching game.
If $G$ is bipartite, then we speak of a {\it $b$-assignment game}. {These games were introduced by \citeA{So92} as multiple partner assignment games.
{If $b\equiv t$ for some integer~$t$, then we speaks of a $t$-matching game or a $t$-assignment game.}
N}ote that ${1}$-assignment games correspond to assignment games.
A $b$-matching game {is} called a {\it multiple partners matching game} {in the paper of \shortciteA{BKPW18}}.

\begin{figure}
\centering
\begin{tikzpicture}[scale=1,rotate=0, blue]
\draw
(0, 1) node[circle, blue, scale=0.8,draw](a){\small $1$}
(1.5, 2) node[label=above:{\small[$2$]},circle, blue, scale=0.8,draw](b){\small  $2$}
(1.5, 0) node[circle, blue, scale=0.8,draw](c){\small $3$}
(3, 3) node[circle, blue, scale=0.8,draw](d){\small $4$}
(3, 1) node[label=right:{\small [$3$]},circle, blue, scale=0.8,draw](e){\small $5$}
(3, -1) node[circle, blue, scale=0.8,draw](f){\small $6$};
\draw[-] (a) -- node[above] {\small $2$} (b);
\draw[-] (b) -- node[above] {\small $2$} (d);
\draw[-] (d) -- node[right] {\small $3$} (e);
\draw[-] (b) -- node[above] {\small $1$} (e);
\draw[-] (a) -- node[below] {\small $1$} (c);
\draw[-] (c) -- node[above] {\small $3$} (e);
\draw[-] (c) -- node[below] {\small $2$} (f);
\draw[-] (e) -- node[right] {\small $2$} (f);
\end{tikzpicture}
\caption{{A $b$-matching game $(N,v)$ with six players, where $b\equiv 1$ except
$b(2)=2$ and $b(5)=3$, so $b^*=3$.
Note that $v(N)=10$ (take $M=\{12,35,45,56\}$).}}\label{f-fourth}
\end{figure}
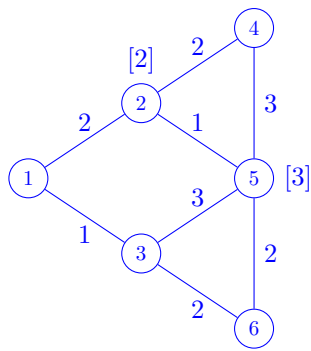

{
In the next definition, the sets $N$ (of players) and $V$ (of vertices) are different, as we now associate each player~$i\in N$ with a distinct subset~$V_i$ of the vertex set~$V$.}

\bigskip
\noindent
{\bf Definition 3.} A {\it partitioned matching game} {\shortcite{BKPP19}} on a graph $G=(V,E)$ with a positive edge weighting~$w$ and partition $(V_1,\ldots,V_{{|N|}})$ of $V$ is the game $(N,v)$, where $N=\{1,\ldots,{|N|}\}$, and for $S\subseteq N$, the value $v(S)$ is the maximum weight $w(M)$ over all matchings $M$ of $G[\bigcup_{i\in S}V_i]$.

\bigskip
\noindent
Partitioned matching games {were introduced as} {\it generalized matching games} by~\shortciteA{BKPP19} {and are also known as}
{\it international matching games}~\cite{KK19}.
We say that $$c=\max\{|V_i| \; |\; 1\leq i\leq |N|\}$$ is the
 {\it width} of $(N,v)$.
Note that if $V_i=\{i\}$ for every $i\in N$, then we obtain a matching game (which has width~$1$).
We refer to Figure~\ref{f-thirdleft} for an example of a partitioned matching game with width~$3$.

\begin{figure}
\centering
{\begin{tikzpicture}[scale=1,rotate=0]
\draw
(0, 1) node[circle, black, scale=0.8,draw](a){\small$1$}
(1.5, 2) node[circle, black,scale=0.8, draw](b){\small$2$}
(1.5, 0) node[circle, black, scale=0.8,draw](c){\small$3$}
(3, 3) node[circle, black, scale=0.8,draw](d){\small$4$}
(3, 1) node[circle, black, scale=0.8,draw](e){\small$5$}
(3, -1) node[circle, black, scale=0.8,draw](f){\small$6$}
(-0.5, 2) node[scale=0.8](h){\small$V_{1}$}
(3.7, 3) node[scale=0.8](i){\small$V_{2}$}
(3.7, -1) node[scale=0.8](j){\small$V_{3}$};
\draw[-] (a) -- node[above] {\small$2$} (b);
\draw[-] (b) -- node[above] {\small$2$} (d);
\draw[-] (d) -- node[right] {\small$3$} (e);
\draw[-] (b) -- node[above] {\small$1$} (e);
\draw[-] (a) -- node[below] {\small$1$} (c);
\draw[-] (c) -- node[below] {\small$3$} (e);
\draw[-] (c) -- node[below] {\small$2$} (f);
\draw[-] (e) -- node[right] {\small$2$} (f);
\draw[dashed] (0.97,1) ellipse (1.3 and 1.5);
\draw[dashed] (3,2) ellipse (0.6 and 1.4);
\draw[dashed] (3,-1) ellipse (0.35 and 0.35);
\end{tikzpicture}}
\caption{{A partitioned matching game $(N,v)$ of width $c=3$ defined on a graph $G=(V,E)$ with an edge weighting $w$. Note that $N$ consists of three players and that $V$ is partitioned into $\{1,2,3\},\{4,5\},\{6\}$, as indicated by the dotted circles. Also note that $v(N)=7$.}}\label{f-thirdleft}
\end{figure}
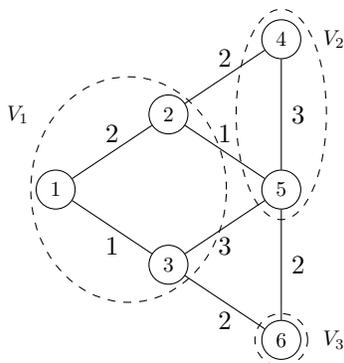

We can now model an international KEP {with pairwise exchanges only} as a partitioned matching game $(N,v)$ that is defined
on a compatibility graph $G=(V,E)$ where
\begin{itemize}
\item the set $N{=\{1,\ldots,|N|\}}$ is a set of countries participating in the international KEP;
\item the set $V$ is partitioned into {subsets} $V_1,\ldots,V_{{|N|}}$, where $V_i$ consists of the patient-donor pairs of country~$i$ for every $i\in {N}$; and
\item the set $E$ may have an edge weighting $w$, where $w(uv)$ expresses the utility of a transplant exchange between patient-donor pairs $u$ and $v$ (if the aim is to maximize the number of transplants we can set $w\equiv 1$).
\end{itemize}
To compute
the target numbers for each of the countries, \shortciteA{BBKP22} used {two solution concepts, the benefit value and the contribution value, which can be computed in linear time,\footnote{{This follows immediately from their definitions, see~\shortciteA{BBKP22}.}} and two solution concepts, the Shapley value and the nucleolus, which in general take exponential time to compute; in particular, they employed the state-of-the-art method~\shortcite{BFN21} for computing the nucleolus.}
From their simulations, the Shapley value yielded the best results. This is in line with the findings of \shortciteA{KNPV20} and \shortciteA{BGKPPV20}, {who considered international KEPs that} allowed cycles of length~$3$, but for simulations involving fewer countries than in the study of \shortciteA{BBKP22} {(see also Section~\ref{s-2}).}

Apart from international KEPs, partitioned matching games can also be used to model
economic settings where multi-organizations own pools of clients~\shortcite{GMP13}.
As we will discuss further in Section~\ref{s-co}, \shortciteA{BKPP19} performed a theoretical study for partitioned matching games and showed how $b$-matching games and partitioned matching games are related to each other.

\medskip
\noindent
We finish this section by defining a known relaxation of matching games. A {\it fractional matching game} on a graph $G=(V,E)$ with a positive edge weighting~$w$ is the game $(N,v)$ where $N=V$ and for $S\subseteq N$, the value $v(S)$ is the maximum weight~$w(M)$ over all fractional matchings $M$ of $G[S]$. We can define the notions of a {\it fractional $b$-assignment game} and {\it fractional $b$-matching game} analogously. These notions have been less studied than their non-fractional counterparts and are not the focus of our survey. Nevertheless, we will mention relevant complexity results for these games as well.

\section{Network Bargaining Games}\label{s-nbg}

\medskip
\noindent
Some economic situations involve only preferences of players. These situations are beyond the scope of this survey.  The economic setting that is described in this section includes payments.
As we will explain in Section~\ref{s-co}, this setting is closely related to the setting of $b$-matching games.

Let $G$ be a graph with a positive edge weighting $w$ and a positive vertex capacity function~$b$. Previously we defined a $b$-matching game $(N,v)$ on $(G,b,w)$, but the triple $(G,b,w)$ is also known as a
{\it network bargaining game}. Again, the vertices are players, but now an edge $ij$ represents
a possible contract between players~$i$ and~$j$.
The weight $w(ij)$ represents the value of the contract, while the {\it capacity} $b(i)$ is the maximum number of contracts player~$i$ can commit to.  A $b$-matching $M$ can now be viewed as a set of pairwise contracts between players. However, $M$ might be {\it blocked} by two players $i$ and $j$. To define the notion of a blocking pair in this context we need some further terminology.

If $ij\in M$, we distribute the value $w(ij)$ of the contract $ij$ by defining {\it pay-offs} $p(i,j)\geq 0$ and $p(j,i)\geq 0$ to $i$ and~$j$, respectively, such that $$p(i,j)+p(j,i)=w(ij).$$
\noindent
If $ij\notin M$, then we set $p(i,j)=p(j,i)=0$. The vector $p$ with entries $p(i,j)$ is a {\it pay-off} and the pair $(M,p)$ is a {\it solution} for $(G,b,w)$.
The {\it total pay-off vector} $p^t$ is defined by $p^t(i)=\sum_{j:ij\in E}p(i,j)$ for every $i\in N$.
The {\it utility} $u_p(i)$ {is equal to $\min\{p(i,j)\; |\; ij\in M\}$ if $i$ is incident to $b(i)$ edges of $M$; otherwise $u_p(i)=0$.}
We can now define a {\it blocking pair} with respect to $(M,p)$ as a pair~$\{i,j\}$ with $ij\in E\setminus M$ such that  $$u_p(i)+u_p(j)<w(ij).$$
\noindent
The goal is to decide if $(G,b,w)$ has a {\it (pairwise) stable} solution, that is, a solution with no blocking pairs. This problem is called the {\sc Stable Fixtures with Payments} (SFP) problem {\shortcite{BKPW18}}.
Restrictions of SFP are called:
\begin{itemize}
\item {\sc Multiple Partners Assignment} (MPA) if $G$ is bipartite {\cite{So92}};
\item {\sc Stable Roommates with Payments} (SRP) if $b\equiv 1$ {\cite{EK01}};
\item {\sc Stable Marriage with Payments} (SMP) if $G$ is bipartite, $b\equiv 1$ {\cite{KB57}}.
\end{itemize}
\noindent
We denote instances of SRP and SMP by $(G,w)$ instead of $(G,1,w)$. 

{It is known that SFP (and thus MPA, SMP, SRP) is polynomial-time solvable~\shortcite{BKPW18,FGK13}.}

The main reason for discussing network bargaining games in our survey is due to their insightful relationship with the core of $b$-matching games (see Section~\ref{s-co}).
{However,} we {also} note the well-known result of \citeA{KT08} who gave a polynomial-time algorithm {for finding a} {\it balanced} stable solution~$(M,p)$ for  
{SRP. Afterwards, \shortciteA{BHIM10} proved that a solution $(M,p)$ for an instance $(G,w)$ of {\sc SRP} is stable and balanced if and only if $p^t$ belongs to the intersection of the core and 
the pre-kernel of the matching game $(N,v)$ defined on $(G,w)$. Balanced stable solutions are beyond the scope of our survey; for the sake of completeness, we include their definition below.} 

{\shortciteA{FGK13} extended the original definition of a balanced stable solution from SPR to SFP.  
Given a stable solution $(M,p)$ for an instance $(G,b,w)$ of SFP, we let $$\alpha_p(i)=\max (0, \max\{w_{ij}-u_p(i)\;|\;  ij\in E(G)\setminus M\})$$ be the {\it outside gain}, which is the maximum that player $i$ can gain on top of 
its total payoff $p^t(i)$) from any neighbour $j$ with $ij\notin M$. A stable solution $(M,p)$ is {\it balanced} if for every $ij\in M$, 
$$p(i,j)-\alpha_p(i) = p(j,i)-\alpha_p(j).$$  Intuitively, this means that the value $w_{ij}$ is shared between $i$ and $j$ in a balanced way: the outside gain of player $i$ is equal to the outside gain of player $j$.
We discuss the paper of \shortciteA{FGK13} in more detail in the next section.}

\section{Complexity Aspects of the Core}\label{s-co}

\medskip
\noindent
In Section~\ref{s-thecore} we survey complexity results for the problems of core membership, core non-emptiness and finding core allocations for matching games and the variants of matching games defined in Section~\ref{s-mg}. In Section~\ref{s-stable} we survey the known results for core stabilizers for matching games.

\subsection{Testing Core Membership, Non-Emptiness and Finding Core Allocations}\label{s-thecore}

\medskip
\noindent
We consider three natural problems, which have also been studied for other games and with respect to other solution concepts. Each of these problems takes as input a game $(N,v)$.
\begin{itemize}
\item [{\bf P1.}] {(core membership) decide} if a given allocation $x$ belongs to the core, or else find a coalition~$S$ with $x(S)<v(S)$;
\item [{\bf P2.}] {(core non-emptiness)} {decide} if the core is non-empty; and
\item [{\bf P3.}] {(core allocation computation)} find an allocation in the core (if it is non-empty).
\end{itemize}
\noindent
We note that if P1 is polynomial-time solvable for some class of games ${\cal G}$, then by using the ellipsoid method~\shortcite{GLS81,Ka79}, it follows that P2 and P3 are also polynomial-time solvable for ${\cal G}$.

\medskip
\noindent
In the remainder we {explain} that the problems SMP, MPA, SRP and SFP are closely related to problems P1--P3 for the corresponding matching game variant. To help the reader keeping track of this we refer to Table~\ref{t-2} for an overview (below we will explain the results in this table as part of our discussion).

\begin{table}
\begin{center}
\small
\begin{tabular}{l|l|l}
& \multicolumn{1}{|c|}{bipartite graphs}  & \multicolumn{1}{|c}{general graphs} \\
 \hline \hline
 $b\equiv 1$    &  \multicolumn{1}{|c|}{SMP}         & \multicolumn{1}{|c}{SRP} \\
 & \multicolumn{1}{|c|}{assignment game} & \multicolumn{1}{|c}{matching game}  \\ \hline
{stable} solution                       & {\it exists}$^{1}$      &{{\it may not exist}}\\
        core               & {\it always non-empty}$^{2}$        &{\it may be empty}\\
\hline \hline
 any $b$   & \multicolumn{1}{|c|}{MPA}   & \multicolumn{1}{|c}{SFP} \\
& \multicolumn{1}{|c|}{$b$-assignment game} & \multicolumn{1}{|c}{$b$-matching game} \\ \hline
           {stable} solution            & {\it exists}$^{3}$    &{{\it may not exist}}\\
core & {\it always non-empty}$^{3}$    &{\it may be empty}\\
\end{tabular}
\caption{The four variants of network bargaining games and their 
counterpart matching games. {Recall that} SFP (and thus MPA, SMP, SRP) is polynomial-time solvable~\shortcite{BKPW18,FGK13}. 
{
The references for the statements on the existence of a stable solution and on the core are $^{1}${\citeA{KB57}},  $^{2}${\citeA{SS72}},  $^{3}${\citeA{So92}}.}
See Table~\ref{t-tabtab} for a summary of the complexity results for matching games.}\label{t-2}
\normalsize
\end{center}
\end{table}

\medskip
\noindent
We start by considering the assignment games and $b$-assignment games. {The problems}
SMP and MPA are closely related to problems~P2 and~P3 for assignment games and $b$-assignment games, respectively.
Namely, \citeA{SS72} proved that every core allocation of an assignment game {corresponds to} a
pay-off vector in a stable solution for SMP and vice versa. Moreover, \citeA{KB57} proved that SMP, and thus P2, only has yes-instances and that a stable solution, and thus a core allocation of every assignment game, can be found in polynomial time. Finally, \citeA{So92} extended all these results to MPA and $b$-assignment games; recall that $p^t$ denotes the total pay-off vector.

\begin{theorem}[\citeR{So92}]\label{t-so92}
Every instance $(G,b,w)$ of {\sc MPA} has a stable solution, which can be found in polynomial time.
Moreover, if $(M,p)$ is a stable solution for $(G,b,w)$, then $M$ is a maximum weight $b$-matching of $G$ and $p^t$ is a core allocation of the $b$-assignment game defined on $(G,b,w)$.
\end{theorem}

\noindent
{The converse of the second statement of Theorem~\ref{t-so92} does not hold, as there may exist an allocation in the core which does not correspond to a stable solution~\cite{So92}.} 

In contrast to the situation for assignment games, the core of a matching game defined on a non-bipartite graph may be empty; just recall the example from Section~\ref{s-mg} where $G$ is the triangle and $w\equiv 1$. Problem~P1 is linear-time solvable for matching games, and thus also for assignment games. Recall from Section~\ref{s-mg} that it suffices to check if for an allocation~$x$, we have for every $ij\in E$,
$$x_i+x_j\geq w(ij).$$
\noindent
As P1 is polynomial-time solvable, we immediately have that P2 and P3 are also polynomial-time solvable for matching games. In particular, the following connection with SRP is well known; note that for SRP, total pay-off vectors and pay-off vectors are in 1-1-correspondence (see also Example~1).

\begin{theorem}[\citeR{EK01}]\label{t-srp}
A solution $(M,p)$ for an instance $(G,w)$ of {\sc SRP} is stable if and only if $M$ is a maximum weight matching of $G$ and $p^t$ is a core allocation of the matching game defined on $(G,w)$.
\end{theorem}

\noindent
\shortciteA{DIN99} followed a more direct approach for solving P2 and~P3 for matching games. Later, a faster algorithm for P2 and P3 was given by \shortciteA{BKP12}. This algorithm
{runs in $O(nm+n^2\log n)$ time and} 
is based on an alternative characterization of the core of a matching game, which follows from a result of \citeA{Ba65}.

\begin{theorem}[\shortciteR{BKP12}]\label{t-sol}
The core of a matching game $(N,v)$, on a graph $G$ with positive edge weighting $w$, is non-empty if and only if the maximum weight of a matching
in $G$ equals the maximum weight of a half-matching in $G$.
\end{theorem}

\noindent
In order to use Theorem~\ref{t-sol} algorithmically, \shortciteA{BKP12} used the duplication technique of \citeA{NT75}, as illustrated in Figure~\ref{f-first}.
That is, for an input graph $G$ with edge weighting~$w$, first construct in polynomial-time the so-called \emph{duplicate} $G^d$ with edge weighting~$w^d$, as follows.
For every vertex $i\in V$, we introduce two vertices~$i'$ and~$i''$ in~$G^d$. For every edge $ij\in E$, we introduce the edges $i'j''$ and $i''j'$, each of weight $w^d(i'j'')=w^d(i''j')=\frac{1}{2}w(ij)$, in $G^d$. It is not difficult to see that the maximum weight of a half-matching in $G$ equals the maximum weight of a matching in $G^d$.
Hence, it remains to compare the latter weight with the maximum weight of a matching in $G$. One can compute both these values in $O({n}^3)$ time by using Edmonds' algorithm~\cite{Ed65}.

\begin{figure}[t]
\begin{minipage}[t]{0.48\linewidth}
\centering
\begin{tikzpicture}[scale=1,rotate=0]

\draw
(0, 1) node[circle, black, scale=0.8,draw](a){\small $1$}
(1.5, 2) node[circle, black, scale=0.8,draw](b){\small $2$}
(1.5, 0) node[circle, black, scale=0.8,draw](c){\small $3$}
(3, 3) node[circle, black, scale=0.8,draw](d){\small $4$}
(3, 1) node[circle, black, scale=0.8,draw](e){\small $5$}
(3, -1) node[circle, black, scale=0.8,draw](f){\small $6$};

\draw[-] (a) -- node[above] {\small $2$} (b);
\draw[-] (b) -- node[above] {\small $2$} (d);
\draw[-] (d) -- node[right] {\small $3$} (e);
\draw[-] (b) -- node[above] {\small $1$} (e);
\draw[-] (a) -- node[below] {\small $1$} (c);
\draw[-] (c) -- node[above] {\small $3$} (e);
\draw[-] (c) -- node[below] {\small $2$} (f);
\draw[-] (e) -- node[right] {\small $2$} (f);
\end{tikzpicture}
\end{minipage}
\begin{minipage}[t]{0.48\linewidth}
\centering
\begin{tikzpicture}[scale=1,rotate=0]
\draw
(0, 2) node[rectangle, black, scale=0.8,draw](a){\small $1^{\prime}$}
(1, 2) node[rectangle, black, scale=0.8,draw](b){\small $2^{\prime}$}
(2, 2) node[rectangle, black, scale=0.8,draw](c){\small $3^{\prime}$}
(3, 2) node[rectangle, black, scale=0.8,draw](d){\small $4^{\prime}$}
(4, 2) node[rectangle, black, scale=0.8,draw](e){\small $5^{\prime}$}
(5, 2) node[rectangle, black, scale=0.8,draw](f){\small $6^{\prime}$}
(0, -1) node[rectangle, black, scale=0.8,draw](g){\small $1^{\prime\prime}$}
(1, -1) node[rectangle, black, scale=0.8,draw](h){\small $2^{\prime\prime}$}
(2, -1) node[rectangle, black, scale=0.8,draw](i){\small $3^{\prime\prime}$}
(3, -1) node[rectangle, black, scale=0.8,draw](j){\small $4^{\prime\prime}$}
(4, -1) node[rectangle, black, scale=0.8,draw](k){\small $5^{\prime\prime}$}
(5, -1) node[rectangle, black, scale=0.8,draw](l){\small $6^{\prime\prime}$};

\draw[-] (a) -- node[right] {}(h);
\draw[-] (a) -- (i);
\draw[-] (b) -- (g);
\draw[-] (b) -- (j);
\draw[-] (b) -- (k);
\draw[-] (c) -- (g);
\draw[-] (c) -- (k);
\draw[-] (c) -- (l);
\draw[-] (d) -- (h);
\draw[-] (d) -- (k);
\draw[-] (e) -- (h);
\draw[-] (e) -- (i);
\draw[-] (e) -- (j);
\draw[-] (e) -- (l);
\draw[-] (f) -- (i);
\draw[-] (f) -- (k);

\end{tikzpicture}
\end{minipage}
\caption{{\it Left:} an example of a matching game on a graph $G$.
{\it Right:} the duplicate~$G^d$ of $G$; note that, for instance, $w^d(3'5'')=\frac{3}{2}$. See also Examples~1 and~2.}\label{f-first}
\end{figure}
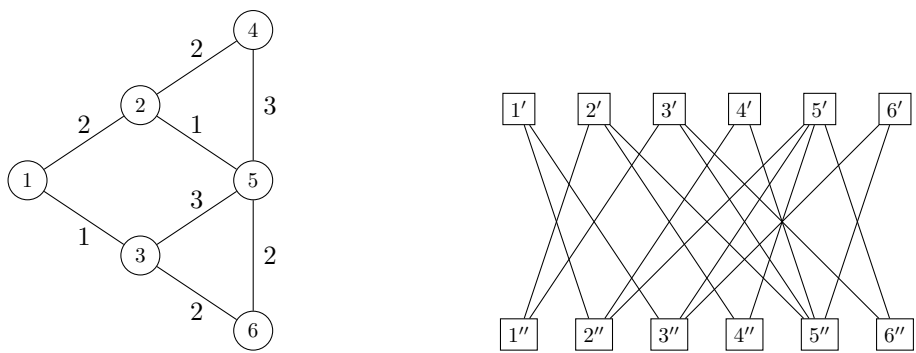

{We now give two examples. The goal of Example~1 is to illustrate the working of Theorems~\ref{t-srp} and~\ref{t-sol}, whereas Example~2 gives another illustration of the power of Theorem~\ref{t-sol}.}

\medskip
\noindent
{\bf Example 1.} Consider the graph~$G$ and its duplicate $G^d$ in Figure~\ref{f-first}. The matching $M=\{12,36,45\}$ is a maximum weight matching of $G$ with maximum weight~$7$. The maximum weight of a matching in $G^d$ is also equal to~$7$. Hence, by Theorem~\ref{t-sol}, we find that the core is non-empty and thus
the nucleolus $\eta=(\frac{1}{2}, \frac{3}{2}, \frac{3}{2}, 1, 2, \frac{1}{2})$ is a core allocation, as can be readily verified.
Moreover, the pair $(M,p)$, where $p$ is defined by $p(1,2)=\frac{1}{2}$, $p(2,1)=\frac{3}{2}$, $p(3,6)=\frac{3}{2}$, $p(4,5)=1$, $p(5,4)=2$ and $p(6,3)= \frac{1}{2}$ and else $p(i,j)=0$, is a stable solution for the corresponding instance $(G,w)$ of SRP, {in line with Theorem~\ref{t-srp}.}
On the other hand, the Shapley value $\phi=
(\frac{23}{30}, \frac{41}{30}, \frac{36}{30}, \frac{31}{30}, \frac{55}{30},\frac{24}{30})$ is not a core allocation, as
$v(\{4,5\})=3>\frac{86}{30}=\phi(\{4,5\})$. \dia

\medskip
\noindent
{\bf Example 2.} Let $G'$ be the graph obtained from the graph $G$ in Figure~\ref{f-first} after removing vertex~6.
Unlike the matching game on $G$, the matching game on $G'$ has an empty core. The reason is that
the maximum weight ($5\frac{1}{2}$) of a matching in~$G'^d$ is larger than the maximum weight~($5$) of a matching in $G'$, and we can apply Theorem~\ref{t-sol} again. \dia

\medskip
\noindent
By Theorem~\ref{t-so92},  every $b$-assignment game is a yes-instance for P2 and P3 is polynomial-time solvable for $b$-assignment games. In contrast, \shortciteA{BKPW18} proved that P1 is {co\NP}-complete even for uniform $3$-assignment games.

On the positive side, \shortciteA{BKPW18}  also showed that for $b\leq 2$, P1 (and thus P2 and P3) is polynomial-time solvable even for $b$-matching games.
They did this by a reduction to the polynomial-time solvable {\it tramp steamer} problem. The latter problem is that of finding a cycle~$C$ in a graph $G$ with maximum profit-to-cost ratio
$$\frac{p(C)}{w(C)},$$ where $p$ and $w$ represent profit and cost functions, respectively, that are defined on $E(G)$.

Before considering partitioned matching games, it remains to discuss P2 and P3 for $b$-matching games for which $b\not\leq 2$. Theorem~\ref{t-srp} implies that besides the aforementioned direct algorithms, any polynomial-time algorithm for SRP can also be used for immediately solving P2 and~P3 for the corresponding matching game. As such, the following theorem for SFP, shown
independently by \shortciteA{BKPW18} and \shortciteA{FGK13}\footnote{The algorithm of \shortciteA{FGK13} even computes a balanced stable solution {by mapping an instance of the problem to an instance of SRP where each vertex~$i$ has as many copies as its original capacity~$b(i)$, and then applying the result of \citeA{KT08} on finding balanced stable solutions for SRP (see Section~\ref{s-nbg}). T}he {$O(n^2m \log (n^2/m))$-time} algorithm of \shortciteA{BKPW18} is faster, as it makes use of an analogue of Theorem~\ref{t-sol} for SFP.}, is initially promising for $b$-matching games.

\begin{theorem}[\shortciteR{BKPW18,FGK13}]\label{t-both} \emph{SFP}  can be solved in polynomial time.
Moreover, if $(M,p)$ is a stable solution for an instance $(G,b,w)$ of {\sc SFP}, then $M$ is a maximum weight $b$-matching of $G$ and $p^t$ is a core allocation of the $b$-matching game defined on $(G,b,w)$.
\end{theorem}

\noindent
By Theorem~\ref{t-both} we can first solve SFP for a given instance $(G,b,w)$ and if a stable solution is found, then we immediately obtain a core allocation for the $b$-matching game defined on $(G,b,w)$.
However, Theorem~\ref{t-both} does not give us anything if $(G,b,w)$ has no stable solutions.
The following example given by~\shortciteA{BKPW18} shows that {the corresponding $b$-matching game may still have a core allocation},
 even if $b\leq 2$. \shortciteA{FGK13} gave a different example {for $b\leq 2$}, where the core allocation even belongs to the pre-kernel. 

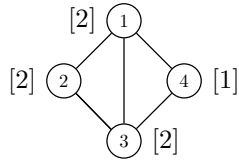
\begin{figure}[t]
\centering
\begin{tikzpicture}[scale=0.8,rotate=0,line width=0.48]
\draw
(0, 0) node[label=left:{\small[$2$]},circle, black, scale=0.7,draw](a){\small $2$}
(1, -1) node[label=right:{\small[$2$]},circle, black, scale=0.7,draw](b){\small $3$}
(2, 0) node[label=right:{\small[$1$]},circle, black, scale=0.7,draw](c){\small $4$}
(1, 1) node[label=left:{\small[$2$]},circle, black, scale=0.7,draw](d){\small $1$};
\draw[-] (a) -- (b)--(c)--(d)--(a);
\draw[-] (a) --(b);
\draw[-] (b) --(d);
\end{tikzpicture}
\caption{The graph $G$ with $b(1)=b(2)=b(3)=2$; $b(4)=1$; and $w\equiv 1$ from Example~3.}\label{f-core}
\end{figure}

\medskip
\noindent
{\bf Example 3.} Consider the graph $G$ {with vertex capacity function $b$ from} Figure~\ref{f-core}. We have $v(N)=3$ in the corresponding uniform $b$-matching game; take $M=\{12,23,34\}$. Note that  $(1,1,1,0)$ is a core allocation. \shortciteA{BKPW18} proved that an instance of SFP has a stable solution if and only if the maximum weight of a $b$-matching in the underlying graph equals the maximum weight of a half-$b$-matching. Now, the maximum weight of a half $b$-matching in $G$ is $3\frac{1}{2}$; take $f(12)=f(23)=1$ and $f(13)=f(14)=f(34)=\frac{1}{2}${, while the maximum weight of a $b$-matching is $v(N)=3$}. So no stable solution for $(G,b,w)$ exists.\dia

\bigskip
\noindent
From Example~3 we find that Theorem~\ref{t-srp} cannot be generalized to SFP.  Indeed, some time later, \shortciteA{BKPP19} proved that P2 and thus P3 are in fact {co\NP}-hard even for uniform $b$-matching games if $b\leq 3$.

\begin{table}
\centering
\begin{tabular}{c|c|c|c}
     & P1 & P2     &P3\\
\hline
assignment game & poly &yes$^{1}$  &poly\\
\hline
matching game   &poly (trivial) &poly$^{2}$  &poly\\
\hline \hline
\multicolumn{4}{c}{$b$-assignment game} \\ \hline 
if $b\leq 2$   & poly  &yes  &poly\\
if $b\leq 3$ 
				 &co\NP c$^{3}$ &yes$^{4}$  &poly$^{4}$ \\		
\hline \hline
\multicolumn{4}{c}{$b$-matching game}\\  \hline
if $b\leq 2$  & poly$^{3}$  &poly      & poly      \\
if $b\leq 3$
             &co\NP c$^{3}$ 			 &co\NP h$^{5}$  & co\NP h \\
\hline \hline
\multicolumn{4}{c}{partitioned  
matching game} \\ \hline
if $c\leq 2$  & poly$^{5}$  &poly$^{5}$      & poly      \\
if $c\leq 3$             &co\NP c$^{5}$			 &co\NP h$^{5}$   & co\NP h \\			
\end{tabular}
\caption{Complexity dichotomies for the core (P1--P3), with the following short-hand notations,
yes: all instances are yes-instances; poly: polynomial-time; co\NP c: {co\NP}-complete; and co\NP h: {co\NP}-hard.
{The seven hardness results in the table hold even if $w\equiv 1$. The references are $^{1}$\citeA{SS72}, $^{2}$\shortciteA{DIN99}, $^{3}${\citeA{BKPW18}}, $^{4}${\citeA{So92}}, $^{5}${\citeA{BKPP19}}.} The unreferenced results follow from the referenced results.}\label{t-tabtab}
\end{table}

All the results for P1--P3 are summarized in Table~\ref{t-tabtab}. This table also displays a dichotomy for partitioned matching games. This dichotomy is obtained after establishing a close relationship between the core of partitioned matching game and the core of a $b$-assignment game, which we discuss below.

\bigskip
\noindent
First, let $(N,v)$ be a partitioned matching game of width~$c$ defined on a graph $G=(V,E)$ with a positive edge weighting $w$ and with a partition
$(V_1, \dots,V_{{|N|}})$ of $V$. We assume that $c \ge 2$, as otherwise we obtain a matching game. Construct a graph $\overline{G}=(\overline{N},\overline{E})$  with a positive vertex capacity function $b$ and a positive edge weighting~$\overline{w}$
as follows.

\begin{itemize}
\item Put the vertices of $V$ into ${\overline N}$ and the edges of $E$ into $\overline{E}$.
\item For each $V_i$, add a new vertex $r_i$ to $\overline{N}$ that is adjacent to all vertices of $V_i$ and to no other vertices in $\overline{G}$.
\item Let $\overline{w}$ be the extension of $w$ to $\overline{E}$ by giving each new edge
weight $2v(N)$.
\item In ${\overline{N}}$, give every $u\in V$, capacity $b(u)=2$ and every $r_i$ capacity $b(r_i)=|V_i|$.
\end{itemize}

\noindent
We denote the $b$-matching game defined on $(\overline{G},b,\overline{w})$ by $(\overline{N},\overline{v})$. See Figure~\ref{f-third} for an example {(based on Figure~\ref{f-thirdleft})}, and note that, by construction, we have $b\leq c$. \shortciteA{BKPP19} used the above construction to show the following theorem.\footnote{In the paper of \shortciteA{BKPP19}, Theorems~\ref{t-t1} and~\ref{t-t2} are explicitly proven for P2. The fact that they hold for P1 and P3 as well is implicit in their proofs{; see 
the work of \shortciteA{BBKPP} for the details.}}

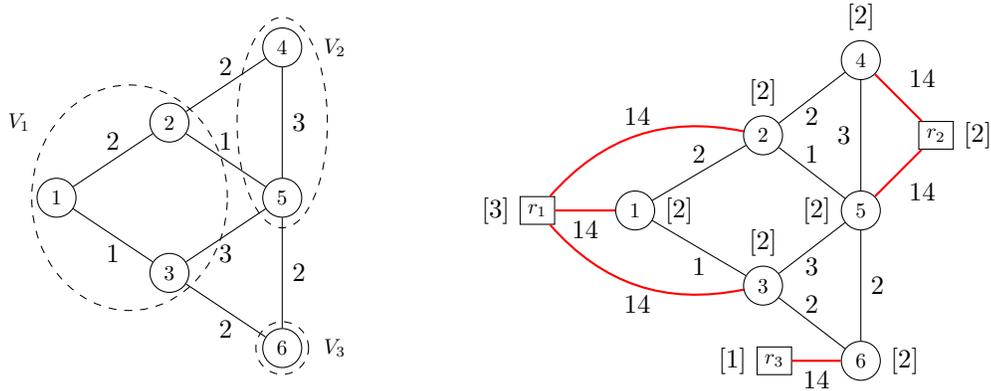
\begin{figure}
\begin{minipage}{0.48\linewidth}
\centering
\begin{tikzpicture}[scale=1,rotate=0]
\draw
(0, 1) node[circle, black, scale=0.8,draw](a){\small$1$}
(1.5, 2) node[circle, black,scale=0.8, draw](b){\small$2$}
(1.5, 0) node[circle, black, scale=0.8,draw](c){\small$3$}
(3, 3) node[circle, black, scale=0.8,draw](d){\small$4$}
(3, 1) node[circle, black, scale=0.8,draw](e){\small$5$}
(3, -1) node[circle, black, scale=0.8,draw](f){\small$6$}
(-0.5, 2) node[scale=0.8](h){\small$V_{1}$}
(3.7, 3) node[scale=0.8](i){\small$V_{2}$}
(3.7, -1) node[scale=0.8](j){\small$V_{3}$};
\draw[-] (a) -- node[above] {\small$2$} (b);
\draw[-] (b) -- node[above] {\small$2$} (d);
\draw[-] (d) -- node[right] {\small$3$} (e);
\draw[-] (b) -- node[above] {\small$1$} (e);
\draw[-] (a) -- node[below] {\small$1$} (c);
\draw[-] (c) -- node[below] {\small$3$} (e);
\draw[-] (c) -- node[below] {\small$2$} (f);
\draw[-] (e) -- node[right] {\small$2$} (f);
\draw[dashed] (0.97,1) ellipse (1.3 and 1.5);
\draw[dashed] (3,2) ellipse (0.6 and 1.4);
\draw[dashed] (3,-1) ellipse (0.35 and 0.35);
\end{tikzpicture}
\end{minipage}
\begin{minipage}{0.48\linewidth}
\centering
\begin{tikzpicture}[scale=1,rotate=0]
\draw
(-1.3, 1) node[label=left:{\small[$3$]},rectangle,black, scale=0.8,draw](g){\small$r_{1}$}
(0, 1) node[label=right:{\small[$2$]},circle, black, scale=0.8,draw](a){\small$1$}
(1.7, 2) node[label=above:{\small[$2$]},circle, black,scale=0.8,draw](b){\small$2$}
(1.7, 0) node[label=above:{\small[$2$]},circle, black, scale=0.8,draw](c){\small$3$}
(3, 3) node[label=above:{\small[$2$]},circle, black, scale=0.8,draw](d){\small$4$}
(3, 1) node[label=left:{\small[$2$]},circle, black, scale=0.8,draw](e){\small$5$}
(4, 2) node[label=right:{\small[$2$]},rectangle,black, scale=0.8,draw](h){\small$r_{2}$}
(1.85, -1) node[label=left:{\small[$1$]},rectangle,black, scale=0.8,draw](i){\small$r_{3}$}
(3, -1) node[label=right:{\small[$2$]},circle, black, scale=0.8,draw](f){\small$6$};
\draw[-] (a) -- node[above] {\small $2$} (b);
\draw[-] (b) -- node[below] {\small$2$} (d);
\draw[-] (d) -- node[left] {\small$3$} (e);
\draw[-] (b) -- node[above] {\small$1$} (e);
\draw[-] (a) -- node[below] {\small$1$} (c);
\draw[-] (c) -- node[below] {\small$3$} (e);
\draw[-] (c) -- node[above] {\small$2$} (f);
\draw[-] (e) -- node[right] {\small$2$} (f);
\path[red][thick] (g) [bend left]edge node[above,black]{\small$14$} (b);
\draw[red][thick] (g) -- node[below,black] {\small$14$} (a);
\path[red][thick] (g) [bend right] edge node[below,black] {\small$14$} (c);
\draw[red][thick] (d) -- node[above right,black] {\small$14$} (h);
\draw[red][thick] (e) -- node[below right,black] {\small$14$} (h);
\draw[red][thick] (i) -- node[below,black] {\small$14$} (f);
\end{tikzpicture}
\end{minipage}
\caption{{\it Left:} {the example from Figure~\ref{f-thirdleft}:} a {partitioned} matching game $(N,v)$ with {three players and with} width $c=3$. {Recall} that $v(N)=7$.
{\it Right:} the reduction to the $b$-matching game $(\overline{N},\overline{v})$. Note that $|\overline{N}|=9$ and that for every $i\in \overline{N}$, $b_i\leq c$.}\label{f-third}
\end{figure}

\begin{theorem}[\shortciteR{BKPP19}]\label{t-t1}
\emph{P1}--\emph{P3} can be reduced in polynomial time from partitioned matching games of width~$c$ to $b$-matching games with {$b\leq c$}.
\end{theorem}

\noindent
As a consequence, P1--P3 are polynomial-time solvable for partitioned matching games of width $c\leq 2$; see also Table~\ref{t-tabtab}.

Now, let $(N,v)$ be a $b$-matching game defined on a graph $G=(V,E)$ with a positive vertex capacity function~$b$ and a positive edge weighting~$w$. Construct a graph $\overline{G}=(\overline{V},\overline{E})$ with a positive edge weighting~$\overline{w}$ and partition~${\cal V}$ of $\overline{V}$ by applying the aforementioned construction of \citeA{Tu54}.

\begin{itemize}
\item Replace each vertex $i \in V$ with capacity $b(i)$ by a set $V_i$ of $b(i)$ vertices $i^1,\ldots,i^{b(i)}$.
\item Replace  each edge $ij \in E$ by a tree $T_{ij}$ connecting the copies of $i$ to the copies of $j$. The tree $T_{ij}$ consists of a central edge with
end-vertices~$i_j$ and~$j_i$, where $i_j$ is adjacent to all copies of $i$, and $j_i$ is adjacent to all copies of $j$.
\item Give every edge in $T_{ij}$ weight $w(ij)$.
\item Let $\overline{V}$ consist of the sets $V_i$ and the $2$-vertex sets $\{i_j,j_i\}$.
\end{itemize}

\noindent
We denote the partitioned matching game defined on $(\overline{G},w)$ with partition~${\cal V}$ by $(\overline{N},\overline{v})$. See Figure~\ref{f-second} (based on Figure~\ref{f-fourth}) for an example.
We let $b^*$ be the maximum $b(i)$-value for the vertex capacity function~$b$.
Note that, by construction, we have that $c=b^*$ and that uniformity is preserved 
{(that is, if the original $b$-matching game was uniform, then the constructed partitioned matching game will be uniform as well)}.
\shortciteA{BKPP19} used the above construction to show the following theorem.

\begin{theorem}[\shortciteR{BKPP19}]\label{t-t2}
\emph{P1}--\emph{P3} can be reduced in polynomial time from $b$-matching games to partitioned matching games of width $c=b^*$, preserving uniformity.
\end{theorem}

\noindent
As a consequence, P1 is {co\NP}-complete and P2, P3 are {co\NP}-hard, even for uniform partitioned matching games of width $c\leq 3$; see also Table~\ref{t-tabtab}.

\begin{figure}
\begin{minipage}{0.48\linewidth}
\centering
\begin{tikzpicture}[scale=1,rotate=0]
\draw
(0, 1) node[circle, black, scale=0.8,draw](a){\small$1$}
(1.5, 2) node[label=above:{\small[$2$]},circle, black, scale=0.8,draw](b){\small$2$}
(1.5, 0) node[circle, black, scale=0.8,draw](c){\small$3$}
(3, 3) node[circle, black, scale=0.8,draw](d){\small$4$}
(3, 1) node[label=right:{\small[$3$]},circle, black, scale=0.8,draw](e){\small$5$}
(3, -1) node[circle, black, scale=0.8,draw](f){\small$6$};
\draw[-] (a) -- node[above] {\small$2$} (b);
\draw[-] (b) -- node[above] {\small$2$} (d);
\draw[-] (d) -- node[right] {\small$3$} (e);
\draw[-] (b) -- node[above] {\small$1$} (e);
\draw[-] (a) -- node[below] {\small$1$} (c);
\draw[-] (c) -- node[above] {\small$3$} (e);
\draw[-] (c) -- node[below] {\small$2$} (f);
\draw[-] (e) -- node[right] {\small$2$} (f);
\end{tikzpicture}
\end{minipage}
\begin{minipage}{0.48\linewidth}
\centering
\begin{tikzpicture}[scale=1,rotate=0]
\draw
(0, 1) node[circle, black, scale=0.8,draw](a){\small$1$}
(0.5,2/3) node[fill,circle, black, scale=0.3,draw](p){}
(1,1/3) node[fill,circle, black, scale=0.3,draw](q){}
(0.5,1+1/3 ) node[fill,circle, black, scale=0.3,draw](g){}
(1,1+2/3 ) node[fill,circle, black, scale=0.3,draw](h){}
(1, 2.45) node(b){\small$V_{2}$}
(1.75,2-1/6) node[circle, black, scale=0.6,draw](x){}
(1.25,2+1/6) node[circle, black, scale=0.6,draw](y){}
(2,2-1/3) node[fill,circle, black, scale=0.3,draw](v){}
(2.5,2-2/3) node[fill,circle, black, scale=0.3,draw](w){}
(2,2+1/3) node[fill,circle, black, scale=0.3,draw](i){}
(2.5,2+2/3) node[fill,circle, black, scale=0.3,draw](j){}
(1.5, 0) node[circle, black, scale=0.8,draw](c){\small$3$}
(2,-1/3) node[fill,circle, black, scale=0.3,draw](r){}
(2.5,-2/3) node[fill,circle, black, scale=0.3,draw](s){}
(2,1/3) node[fill,circle, black, scale=0.3,draw](t){}
(2.5,2/3) node[fill,circle, black, scale=0.3,draw](u){}
(3, 3) node[circle, black, scale=0.8,draw](d){\small$4$}
(3,1+4/3) node[fill,circle, black, scale=0.3,draw](l){}
(3,1+2/3) node[fill,circle, black, scale=0.3,draw](m){}
(3.8, 1) node(e){\small$V_{5}$}
(3.25,1+1/6) node[circle, black, scale=0.6,draw](z){}
(3.25,5/6) node[circle, black, scale=0.6,draw](aa){}
(2.75,1) node[circle, black, scale=0.6,draw](bb){}
(3,-1+4/3) node[fill,circle, black, scale=0.3,draw](n){}
(3,-1+2/3) node[fill,circle, black, scale=0.3,draw](o){}
(3, -1) node[circle, black, scale=0.8,draw](f){\small$6$};
\draw[-] (y) to[out=0,in=90] (v);
\draw[-] (a) -- (g)-- (h)-- (y)-- (i)-- (j)-- (d)-- (l)-- (m)-- (bb)-- (n)-- (o)-- (f)-- (s)-- (r)-- (c)-- (q)-- (p)-- (a);
\draw[-] (c) -- (t)-- (u)-- (bb);
\draw[-] (x) -- (v)-- (w)-- (bb);
\draw[-] (h) -- (x)-- (i);
\draw[-] (m) -- (z);
\draw[-] (m) to[out=-30,in=30] (aa);
\draw[-] (w) -- (z);
\draw[-] (w) to[out=-25,in=145] (aa);
\draw[-] (u) to[out=25,in=-145] (z);
\draw[-] (u) -- (aa);
\draw[-] (n) to[out=30,in=-30] (z);
\draw[-] (n) -- (aa);
\draw[dashed] (3.1,1) ellipse (0.5 and 0.35);
\draw[dashed] (1.5,2) ellipse (0.5 and 0.38);
\draw[dashed] (0,1) ellipse (0.4 and 0.4) node[below,yshift=-.4cm, xshift=-.2cm]{\scriptsize $V_{1}$};
\draw[dashed,rotate around={145:(2.25,-0.5)}] (2.25,-0.5) ellipse (0.5 and 0.35) node[ left,xshift=-.2cm,yshift=-0.4cm]{\scriptsize $\{3_6,6_3\}$};
\end{tikzpicture}
\end{minipage}
\caption{{\it Left}: {the example from Figure~\ref{f-fourth}:} a $b$-matching game $(N,v)$ with six players, where $b\equiv 1$ apart from
$b(2)=2$ and $b(5)=3$, so $b^*=3$.
{Recall} that $v(N)=10$ (take $M=\{12,35,45,56\}$).
{\it Right:} the reduction to the partitioned matching game~$(\overline{N},\overline{v})$. Note that
$|\overline{N}|={14}$ and $c=b^*$.}\label{f-second}
\end{figure}
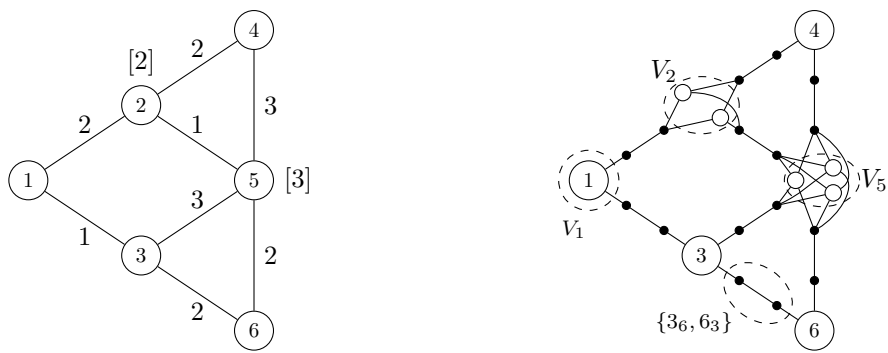

\subsection{Core Stabilizers}\label{s-stable}

\medskip
\noindent
Apart from considering alternative solution concepts, other directions that try to overcome core emptiness have been explored as well. For example, we may ask for the minimum subsidy $\alpha$ such that the adjusted game $(N,v')$ with $v'(N)=v(N)+\alpha$ and
$v'(S)=v(S)$ for $S\subsetneq N$ has a non-empty core~\shortcite{BEMMPRRZ18}.
Recall {from Section~\ref{s-mg}} that the core of a matching game~$(N,v)$ defined on a graph $G=(V,E)$ with some positive edge weighing $w$ is described by all imputations~$x$ with $x_i+x_j\geq w(ij)$ for every $ij\in E$ {\shortcite{DIN99}}. Hence, for matching pairs the minimum subsidy can be computed by solving the following linear program:
\[ \begin{array}{lllcccll}
      &&&\min \alpha=\displaystyle\sum_{i\in N} y_i\\[15pt]
       &&\mbox{s.t.} &x_i+x_j+y_i+y_j &\geq &w(ij) &&\mbox{for all}\; ij\in E\\
       &&            &x(N)           &=    &v(N)\\
       &&            &x_i         &\geq &0     &&\mbox{for all}\; i\in N\\
       &&            &y_i 	    &\geq &0     && \mbox{for all}\; i\in N.
\end{array} \]
Solving this linear program takes polynomial time by using the ellipsoid method as observed by \shortciteA{BKP12}, who also gave a combinatorial algorithm\footnote{{We use the notion of a combinatorial algorithm to indicate that it is an algorithm that does not rely on the ellipsoid method.}} {that runs in $O(nm+n^2\log n)$ time}.

Alternatively, we may want to minimize the total blocking value. \shortciteA{BKP12} showed that the difference between the minimum total blocking value and the minimum subsidy can be arbitrarily large. To solve the problem of finding the minimum total blocking value, they formulated another linear program:

\[ \begin{array}{lllcccll}
  \mbox{(BV)} &&&\min \displaystyle\sum_{ij\in E} z_{ij}\\[15pt]
       &&\mbox{s.t.} &x_i+x_j+z_{ij} &\geq &w(ij) &&\mbox{for all}\; ij\in E\\
       &&            &x(N)           &=    &v(N)\\
       &&            &x_{i}         &\geq &0     &&\mbox{for all}\; ij\in N\\
       &&            &z_{ij} 	    &\geq &0     &&\mbox{for all}\; ij\in E.
\end{array} \]
Solving (BV) takes polynomial time, again using the ellipsoid method, which led \shortciteA{BKP12} to ask the following question:

\begin{open}\label{o-blocking}
Is it possible to also give a polynomial-time combinatorial algorithm for computing the minimum total blocking value for a matching game with an empty core?
\end{open}

\noindent
On the negative side, \shortciteA{BKP12} proved that finding the minimum {\it number} of blocking pairs is \NP-hard even for uniform matching games. However, \shortciteA{KLS15} gave a polynomial-time algorithm
for finding the minimum number of blocking pairs approximately for uniform matching games on sparse graphs {up to a constant factor depending on the density of the graph}.

The following example shows some shortcomings of the above concepts and illustrates that a more robust concept of a stabilizer is needed.

\bigskip
\noindent
{\bf Example 4.} Consider the uniform matching game $(N,v)$ on the graph $K_2+2K_3$, the disjoint union of {a graph consisting of two vertices joined by an edge} and two triangles; see also Figure~\ref{f-k}.
Then, $v(N)=3$ and $(N,v)$ has an empty core. Now, the minimum subsidy, minimum total blocking value and minimum {number of} blocking pairs are all equal to~$1$: set $x\equiv \frac{1}{2}$ on the two triangles and $x\equiv 0$ on the edge.
 However, if we remove the isolated edge from $G$, then the core stays empty. \dia

 \begin{figure}[t]
\centering
\begin{tikzpicture}[scale=0.6,rotate=0,line width=0.48]
\draw
(0, 1) node[circle, black, scale=0.6,draw](a){}
(2, 1) node[circle, black, scale=0.6,draw](b){}
(4, 0) node[circle, black, scale=0.6,draw](c){}
(6, 0) node[circle, black, scale=0.6,draw](d){}
(5, 2) node[circle, black, scale=0.6,draw](e){}
(8.5, 0) node[circle, black, scale=0.6,draw](f){}
(10.5, 0) node[circle, black, scale=0.6,draw](g){}
(9.5, 2) node[circle, black, scale=0.6,draw](h){};

\draw[-] (a) -- (b);
\draw[-] (c) --(d) --(e)--(c);
\draw[-] (f) --(g) --(h)--(f);
\end{tikzpicture}
\caption{The graph $K_2+2K_3$ used in Examples~4 and~5.}\label{f-k}
\end{figure}
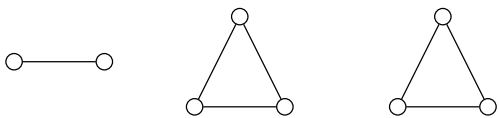

\bigskip
\noindent
We now describe a way to circumvent the issue in Example~4.
Let  $\pi$ be some graph operation, such as the vertex deletion~\vd, edge deletion~\ed, vertex addition~\va\ or edge addition~\ea.
Let $(N,v)$ be a game defined on a graph $G$ with a positive edge weighting~$w$.
Then a {\it $\pi$-stabilizer} for $(N,v)$ is a sequence of operations of type~$\pi$ that transforms $G$ into a graph~$G'$, such that the resulting game has a non-empty core. A set of vertices or edges of $G$ is a {\it blocking set} if its removal from $G$ results in a game with a non-empty core.

\medskip
\noindent
{\bf Example 5.} Let $\pi=\mbox{\ed}$ in Example~4. Then the sequence that deletes one edge of each of the two triangles of $K_2+2K_3$ is a $\pi$-stabilizer of minimum length for the uniform matching game defined on $K_2+2K_3$. Note that every pair of such edges is a blocking set. \dia

\medskip
\noindent
Let $\pi$-{\sc Stabilizer} be the problem of determining a $\pi$-stabilizer of minimum length. This type of problems falls under the classical framework of {\it graph modification}.
For an in-depth survey that includes a wide range of aspects of stabilizers, we refer to the survey of \citeA{Ch17}. Below we give the main complexity results including some new results.

The first result was given by \shortciteA{BBGKP14}
(using the terminology from Section~\ref{s-nbg}).
They proved that
 \ed-{\sc Stabilizer} is \NP-hard for matching games and asked if it is also \NP-hard for uniform matching games. Afterwards, \shortciteA{BCKPS15} gave an affirmative answer to this question even for uniform matching games on {\it factor-critical} graphs (graphs that have a perfect matching after deleting an arbitrary vertex).
In contrast, \shortciteA{CGKPSW19} proved that the variant that minimizes the total weight that might be added to the existing edges is polynomial-time solvable for uniform matching games on factor-critical graphs.

\shortciteA{AHS18} and \shortciteA{IKKKO17} independently proved that {\sc \vd-Stabilizer} is polynomial-time solvable for uniform matching games. \shortciteA{IKKKO17} also proved that {\sc \ea-Stabilizer} and {\sc \va-Stabilizer} are polynomial-time solvable for uniform matching games. The same authors considered the variant where non-edges are given some positive weight and we must find a minimum weight set of non-edges whose addition to the graph results in a game with a non-empty core. They proved that this problem is \NP-hard for arbitrary matching games.
More recently, \citeA{KS20} generalized the positive results of \shortciteA{AHS18} and \shortciteA{IKKKO17} by giving the first polynomial-time algorithm for $\pi$-{\sc Stabilizer} on arbitrary matching games by doing this for $\pi=\mbox{\vd}$.
See Table~\ref{t-tabtab2} for a general overview.

\begin{table}[t]
\centering
\begin{tabular}{l|c|c}
\multicolumn{1}{c|}{$\pi$}    & uniform weights & arbitrary weights\\
\hline
edge deletion {(\ed)}  & {\NP-hard$^{1}$} & {\NP-hard$^{2}$} \\
vertex deletion {(\vd)}  & {polynomial-time solvable$^{3,4}$} & {polynomial-time solvable$^{5}$} \\
edge addition {(\ea)}  & {polynomial-time solvable$^{4}$} & - \\
vertex addition {(\va)}  & {polynomial-time solvable$^{4}$} & - \\
\end{tabular}
\caption{The complexity of $\pi$-{\sc Stabilizer} for matching games. The cases $\pi=\ea$  and $\pi=\va$ {are not well defined and} have not been studied for (arbitrary) matching games. {The references are $^{1}$\shortciteA{BCKPS15}, $^{2}$\shortciteA{BBGKP14}, $^{3}$\shortciteA{AHS18}, $^{4}$\shortciteA{IKKKO17}, $^{5}$\citeA{KS20}.}}\label{t-tabtab2}
\end{table}

\begin{open}
Is it possible to generalize the positive results from Table~\ref{t-tabtab2} to (uniform) $b$-matching games?
\end{open}

\noindent
We finish this section with the following two problems.  First, the  {\sc Minimum Edge Weight Reduction} is to determine a blocking edge set of minimum weight. \shortciteA{BBGKP14} proved that {\sc Minimum Edge Weight Reduction} is \NP-hard for matching games
 (using the terminology of Section~\ref{s-nbg}) and asked the following question, which was also asked by \citeA{Ch17}:

\begin{open}
Determine the computational complexity of {\sc Minimum Edge Weight Reduction} for uniform matching games.
\end{open}

\noindent
Second, the {\sc Minimum Vertex Weight Reduction} is the variant where every vertex~$i$ is given some positive weight $w(i)$ and we must find a blocking vertex set of minimum weight.
 \shortciteA{AHS18} and \shortciteA{IKKKO17} both showed that this problem is \NP-hard for uniform matching games even if there are only two distinct vertex weights. \shortciteA{AHS18} also
showed that the complementary variant that maximizes the weight of the remaining players is \NP-hard.

\section{Complexity Aspects of the Nucleolus}

\medskip
\noindent
The sequence $(\mbox{LP}_r)$ has length at most $n$ (as each time the dimension of the feasible regions decreases). Nevertheless, computing the nucleolus takes exponential time in general due to the computation of the sets ${\cal S}_i$ and the exponential number of inequalities in each linear program of the scheme of \shortciteA{MPS79}.
As shown below, this situation changes for matching games. Note that the nucleolus exists even for $b$-matching games and partitioned matching games, as the set of imputations for these games is non-empty.

We first observe that the least core of matching games with a non-empty core (such as assignment games) has a compact description: for these games, we may restrict the set of constraints to only those for coalitions of size at 
most~$2$.\footnote{{If $x$ is an allocation in this compact description, then $x(S)\geq \sum_{ij\in M}(x_i+x_j)\geq w(M)+\varepsilon=v(S)+\varepsilon$ for any coalition $S$ (where $M$ is a maximum weight matching of $G[S]$), and hence, the compact description describes the least core.}}.
This immediately leads to a polynomial-time algorithm {for the nucleolus} using the ellipsoid method and the scheme of \shortciteA{MPS79}. \citeA{SR94} gave a faster algorithm{, running in $O(n^4)$ time}, for computing the nucleolus of assignment games. By using again the duplication technique of \citeA{NT75},  \shortciteA{BKP12} translated this algorithm into an {$O(n^4)$-time} algorithm for matching games with a non-empty core.

\shortciteA{FKFH98} proved the following result for the nucleon of a matching game; the {definition of the {\it nucleon} is obtained from the definition of the nucleolus by replacing}
(additive) excesses by multiplicative excesses
$e'(S,x)=x(S)/v(S)$ in the definition of the nucleolus. {In this result, the {\it size} of $w$ is the number of bits in its binary representation.}

\begin{theorem}[\shortciteR{FKFH98}]\label{t-faigle}
The {\it nucleon} can be computed in time polynomial in $n$ and the size of~$w$ for matching games.
\end{theorem}

\noindent
\shortciteA{FKFH98} naturally asked if the same holds for the nucleolus. This led to a series of papers, which we discuss below.

\shortciteA{CLZ12} proved that the nucleolus of a fractional matching game can be found in polynomial  time.
\citeA{KP03} gave a polynomial-time algorithm
for uniform matching games; in fact this was shown for matching games with edge weights~$w(ij)$ that can be expressed as the sum of positive vertex weights $w'(i)+w'(j)$ of the underlying graph~$G$~\cite{Pa01}. Afterwards, \citeA{Fa15} extended this result by relaxing the vertex weight condition.
By combining techniques of \citeA{KP03} and \shortciteA{BKP12}, \citeA{Ha17} gave {an $O(n^4)$-time} 
combinatorial
algorithm for computing the nucleolus of a uniform matching game that does not rely on the ellipsoid method.
Finally, \shortciteA{KPT20} solved the open problem of \shortciteA{FKFH98}.

\begin{theorem}[\shortciteR{KPT20}]\label{t-nuc}
The nucleolus can be computed in polynomial time for matching games.
\end{theorem}
The proof of Theorem~\ref{t-nuc} is highly involved. For a least core allocation~$x$, a matching $M$ is {\it $x$-tight} if  the incident vertices of $M$ form a set $S$ with $x(S)=v(S)+\varepsilon_1$  {(so $(x,\varepsilon)$ is an optimal solution for the least core, see also   Section~\ref{s-cgt})}. A matching $M$ is {\it universal} if it is $x$-tight for every least core allocation~$x$. A least core allocation $x$ is {\it universal} if the set of $x$-tight matchings is precisely the set of universal matchings; in particular the nucleolus is a universal allocation. To prove Theorem~\ref{t-nuc}, \shortciteA{KPT20} first characterized universal matchings using classical descriptions of matching polyhedra. They used this characterization to obtain a new description of the set of universal allocations. They showed that this description is of a highly symmetric nature with respect to the least core allocations. This symmetric nature enabled them to give a new description of the least core that is sufficiently compact for computing the nucleolus in polynomial time, by using the aforementioned scheme of \shortciteA{MPS79} and the ellipsoid method.

As the proof of Theorem~\ref{t-nuc} relies on the ellipsoid method, \shortciteA{KPT20} asked the following question.

\begin{open}\label{o-nucleolus}
Is it possible to give a polynomial-time combinatorial algorithm for computing the nucleolus of a matching game?
\end{open}

\noindent
In contrast to the (least) core of assignment games, which is much better understood (see, for example,~\citeR{SR94,Va22b,Wa06}), this requires new structural insights into the least cores of matching games. Progress in this direction might also help in answering an old question of \shortciteA{FKFH98} who asked the same question as \shortciteA{KPT20} but for the nucleon.

\begin{open}\label{o-nucleon}
Is it possible to give a polynomial-time combinatorial algorithm for computing the nucleon of a matching game?
\end{open}

\noindent
The aforementioned {$O(n^4)$-time} algorithm of  \citeA{Ha17} for computing the nucleolus of uniform matching games is a promising start towards resolving Open Problems~\ref{o-nucleolus} and~\ref{o-nucleon}.

\medskip
\noindent
We now turn to $b$-matching games.
 \shortciteA{KTZ21} proved that computing the nucleolus is \NP-hard for uniform $b$-assignment games with $b\leq 3$.
The same authors proved that the nucleolus can be found in polynomial time for two subclasses of $b$-assignment games with $b\leq 2$, one of which is the natural case where $b\equiv 2${,}
and the other one is where {$b\equiv 2$ on one bipartition class}, but $b\equiv 2$ on {only} a subset of vertices of constant size {in the other bipartition class of the underlying graph}.
These results complement a result of \shortciteA{BHIM10} who showed that the nucleolus of a $b$-assignment game can be found in polynomial-time if $b\equiv 1$ on one of the two bipartition classes of the underlying bipartite graph.
Finally, \citeA{KT20} gave a polynomial-time algorithm for computing the nucleolus of
uniform $b$-matching games on graphs of bounded treewidth.

The following problems are still open.

\begin{open}
Determine the complexity of computing the nucleolus of
\begin{itemize}
\item $b$-matching games with $b\leq 2$;
\item fractional $b$-assignment games;
\item fractional $b$-matching games; and
\item {partitioned matching games.}
\end{itemize}
\end{open}

\begin{open}\label{o-bn}
Determine the complexity of computing the nucleon of $b$-matching games {and partitioned matching games}.
\end{open}

\noindent
With respect to Open Problem~\ref{o-bn} we recall that the case $b\equiv 1$ is polynomial-time solvable by Theorem~\ref{t-faigle}.
We also note that the nucleon lies in the {\it multiplicative $\varepsilon$-core}, or equivalently, the {\it $\alpha$-approximate core}, which relaxes the core constraints by replacing $x(S)\geq v(S)$ by $x(S)\geq \alpha v(S)$. \citeA{Va22} proved that every matching game has a non-empty $\frac{2}{3}$-approximate core and that $\alpha=\frac{2}{3}$ is best possible. This result was extended to $b$-matching games by \shortciteA{XLF}. Polynomial-time algorithms for finding a $\frac{2}{3}$-approximate core allocation were given in both papers, but these allocations might not be the nucleon.

\section{Complexity Aspects of the Shapley Value}\label{s-sv}

\medskip
\noindent
Only a few complexity results are known for computing the Shapley value of matching games. \citeA{AK14} proved the following:

\begin{theorem}[\citeR{AK14}]
Computing the Shapley value is \#P-complete for uniform matching games.
\end{theorem}

\noindent
\citeA{AK14} also proved polynomial-time solvability for uniform matching games on graphs $G$ if $G$ has maximum degree at most~$2$ or if $G$ allows a modular decomposition into $k$ cliques for some constant~$k$. They also proved the same results for matching games on graphs $G$ with a positive edge weighting~$w$ if $G$ allows a modular decomposition into $k$ 
{independent sets (sets consisting of pairwise non-adjacent vertices)}; for example, if $G$ is a complete $k$-partite graph. They posed the following problem.

\begin{open}
Determine the computational complexity of finding the Shapley value for fractional matching games.
\end{open}

\noindent
Moreover, \citeA{AK14} asked if there exists a (non-trivial) class of graphs that have at least one vertex of degree at
least~$3$ for which the Shapley value of the corresponding matching game can be computed in polynomial time. This question was answered in the affirmative by \citeA{Bo15} who proved that the Shapley value can be computed in polynomial time for uniform matching games defined on trees.

\citeA{Bo15} asked the natural question whether his result for uniform matching games could be generalized from trees to graph classes of bounded treewidth. \shortciteA{GLS20} showed that this was indeed possible.

Finally, \citeA{KM20} considered $b$-matching games with an additional property. Namely, a game $(N,v)$ is {\it convex} if for all $S,T\subseteq N$, it holds that $v(S)+v(T)\leq v(S\cup T)+v(S\cap T)$.
It is well known that a convex game has a non-empty core that contains the Shapley value~\cite{Sh71}.
Recall from Section~\ref{s-mg} that even uniform matching games with three players may have an empty core. Consequently, not all $b$-matching games are convex. \citeA{KM20} gave {an $O(n\log n+m\alpha(n))$-time algorithm (where $\alpha$ denotes the inverse Ackermann function)} for computing the Shapley value for convex $b$-matching games.

\section{Conclusions}\label{s-con}

\medskip
\noindent
In this survey we discussed computational complexity results on classical solution concepts for matching games, $b$-matching games and partitioned matching games. We also considered the fractional variant of these games whenever relevant. In this final section, we briefly mention two other concepts related to matching games. These concepts are well studied in the literature but beyond the immediate scope of our survey.

\subsection{Hypergraph Matching Games}

\medskip
\noindent
A {\it hypergraph} $H=(V,E)$ consists of a set of vertices $V$ and a set of hyperedges $E$, which are non-empty subsets of $V$. If every edge in $E$ has size~$2$, then~$H$ is a graph. A subset of edges $M\subseteq E$ is a {\it matching} in $H$ if the edges of $M$ are pairwise disjoint. For a subset $S\subseteq V$, we let $H[S]$ denote the hypergraph with vertex set $S$ and
edge set $\{e\; |\; e\in E\; \mbox{and}\; e\subseteq S\}$.

A {\it hypergraph matching game} on a hypergraph~$H=(V,E)$ with a positive edge weighting~$w$ is the game $(N,v)$
where $N=V$ and for $S\subseteq N$, the value~$v(S)$ is the maximum weight~$w(M)$ over all matchings $M$ of $H[S]$. Hypergraph matching games
were introduced by \citeA{KW82} under the name of {\it partitioning games} and
are also known as {\it packing games}~\shortcite{DIN99} or {\it synergy coalition group representation}~\cite{CS06}.

\citeA{CS06} proved that deciding core non-emptiness is \NP-hard for hypergraph matching games. As even computing a value $v(S)$ is \NP-hard for hypergraph matching games, \citeA{KM20b} considered convex hypergraph matching games. They first proved that convex hypergraph matching games can be recognized in polynomial time. They then proved that  the Shapley value of convex hypergraph matching games can be found in polynomial time.

\subsection{NTU Matching Games}

\medskip
\noindent
Cooperative games $(N,v)$ are also known as {\it transferable utility games} (TU games), as it is possible to distribute the value $v(S)$ among the players of a coalition~$S$ in any possible way. {\it Non-transferable utility games} (NTU games) can be seen as a generalization of TU games and play a role in economic settings where goods are indivisible. That is, in an NTU game $(N,v)$, the value function~$v$ now
also associates each non-empty set $S\subseteq N$ with a non-empty set $X(S)$ that consists of vectors $x\in \R^S$. Note that we obtain the corresponding TU game if for each $S\subseteq N$ we have that $X(S)$ consists of all possible positive vectors~$x$ such that $\sum_{i\in S}x_i=v(S)$.
We refer to the work of \citeA{BF16} for a survey on results on NTU matching games and some natural open problems for these games and to a more recent paper of~\citeA{KK19} for complexity results for NTU partitioned matching games, which they call international matching games.

{Furthermore, there is a large body of related work on core stabilizers in this setting; for example on
the problem of minimizing the number of blocking pairs \shortcite{ABM05,BMM10}, 
the problem of maximizing the number of vertices such that these pairs are stable among themselves (in our terminology, the $\vd$-{\sc Stabilizer} problem)~\cite{Ta90},  and on the topic of
manipulative attacks \shortcite{BBHN21,CSS21,EGNRWY23}.}

\begin{acks}
{Benedek acknowledges the support of the National Research, Development and Innovation Office of Hungary (OTKA Grant No.\ K138945). Bir\'o acknowledges the support of the Hungarian Scientific Research Fund (OTKA, Grant No.\ K143858) and the Hungarian Academy of Sciences (Momentum Grant No. LP2021-2). Paulusma acknowledges the support of the Leverhulme Trust (Grant RF-2022-607). All authors thank three anonymous reviewers for their helpful comments on our paper.}
\end{acks}

\vskip 0.2in
\bibliographystyle{theapa}

\end{document}